# Probing orbital currents through inverse orbital Hall and Rashba effects

E. Santos[1], J. L. Costa[1], R.L. Rodríguez-Suárez[2], J. B. S. Mendes[3], and A. Azevedo[1]

[1] *Departamento de Física, Universidade Federal de Pernambuco, Recife, Pernambuco 50670-901, Brazil.*

[2] *Facultad de Física, Pontifícia Universidad Católica de Chile, Casilla 306, Santiago, Chile.*

[3] *Departamento de Física, Universidade Federal de Viçosa, 36570-900 Viçosa, Minas Gerais, Brazil*

**Abstract**

We report a comprehensive experimental investigation of orbital-to-charge conversion in metallic and semiconductor materials, emphasizing the fundamental roles of the inverse orbital Hall effect (IOHE) and the inverse orbital Rashba effect. Using spin pumping driven by ferromagnetic resonance (SP-FMR) and the spin Seebeck effect (SSE), we demonstrate efficient orbital current generation and detection in YIG/Pt/NM structures, where NM is either a metal or a semiconductor. A central finding is the dominance of orbital contributions over spin-related effects, even in systems with weak spin-orbit coupling. In particular, a large enhancement of the SP-FMR and SSE signals is observed in the presence of naturally oxidized Cu in different heterostructures. Furthermore, we identify positive and negative IOHE signals in Ti and Ge, respectively, and extract orbital diffusion lengths in both systems using a diffusive model. Our results confirm the presence of orbital transport and offer valuable insights that may guide the further development of orbitronics.

# 1. INTRODUCTION

Spintronics has revolutionized modern electronics by exploiting the electron spin degree of freedom to process and transport information [1-8]. Many of its key effects, however, rely on strong spin-orbit coupling (SOC) [5-12]. This relativistic interaction links spin ($\hat{S}$) and orbital ($\hat{L}$) angular momenta of the electron, and is described by the Hamiltonian $\lambda \hat{S} \cdot \hat{L}$, where $\lambda$ is a material-dependent parameter that typically increases with the atomic number Z of the constituent elements [13,14]. More recently, orbitronics has emerged as a promising frontier, where the orbital angular momentum (OAM) of electrons is used to generate, manipulate, and detect orbital currents [13,15-19]. Unlike spintronics, orbitronic phenomena can occur even in the absence of SOC, broadening the range of applicable materials to include light metals [13,15,16-21].

Fundamental spintronic effects like the spin Hall effect (SHE) [5,22,23], spin Rashba effect (SRE) [24-27], and spin pumping effect (SP) [28,29], have orbital counterparts. The most prominent is the orbital Hall effect (OHE), where a longitudinal electric field generates a transverse flow of OAM [15,20,30]. This phenomenon arises from orbital textures in momentum space and the intrinsic OAM of electronic states [20]. Similarly, the orbital Rashba effect (ORE) [31-34] arises when inversion symmetry is broken, allowing a charge current to generate a nonequilibrium orbital accumulation, understood as a consequence of chiral OAM textures in reciprocal space. The inverse orbital Hall effect (IOHE) and the inverse orbital Rashba effect (IORE) describe the conversion of orbital currents into charge current, providing a means for their experimental detection [35-44]. The direct and inverse orbital effects are illustrated in Figure 1.

Recent theoretical and experimental studies have reported OHE, ORE, and their inverse effects in 3d, 4d, and 5d transition metals, two-dimensional materials such as transition metal dichalcogenides (TMDs), and topological materials [41,45-47]. In many of these systems, exceptionally large orbital Hall conductivity ($\sigma_{OH}$) has been predicted [48-51] (Figure 2 (a)), and measured [41] (Figure 2 (b)), in most cases surpassing the spin Hall conductivity ($\sigma_{SH}$) of heavy metals with strong SOC. Such large $\sigma_{OH}$ enhances orbital torques and OAM-assisted magnetic switching, opening a pathway toward orbital-based memory and logic devices for next-generation nanoelectronics.

Despite the potential of orbitronics for low-power applications and broader material applicability, a major challenge lies in disentangling spin and orbital angular momentum [52,53]. A further obstacle is the efficient injection of OAM into conducting, ferromagnetic, or two-dimensional systems, which currently limits device implementation [54]. Although orbital currents can exist without spin-orbit coupling, materials with strong SOC provide effective channels for their generation, accumulation, and detection through partial conversion between orbital and spin currents [35-44]. This coupling is particularly pronounced in heavy transition metals such as Pt, where *LS* entanglement allows spin and orbital currents to serve as mutual probes, enabling orbital properties to be extracted from spintronic responses, and vice versa.

In this work, we demonstrate an efficient method for orbital current injection, focusing on the inverse effects (IOHE and IORE), illustrated schematically in Figure 1. We also employ a theoretical model that considers the diffusive behavior of spin and orbital angular momentum, allowing us to obtain several relevant physical parameters. Our experimental results reveal large orbital contributions, exceeding those of spin, and pave the way for disentangling the spin and orbital channels in orbital charge conversion phenomena.

## 2. METHODS

We fabricated a series of heterostructures, including YIG/Pt(2), YIG/Pt(2)/CuO$_x$(3), YIG/Ti(4), and YIG/Ti(4)/CuO$_x$(3), where the number in parentheses represents the film thickness in nanometers and CuO$_x$ represents Cu naturally oxidized over two days. These samples were designed to investigate the influence of oxidation on the orbital Rashba effect at Cu-based interfaces. We also fabricated SiO$_2$/FM/CuO$_x$(3) samples, where FM represents Py (Ni$_{81}$Fe$_{19}$), Co, or Ni. A Py/Pt/CuO$_x$(3) heterostructure was also employed to study the orbital Rashba effect in CuO$_x$, allowing direct comparison with the YIG-based samples. In parallel, we prepared additional heterostructures to address orbital injection and diffusion effects beyond Cu-based interfaces.

To investigate orbital injection into Ti and Ge thin films, we fabricated YIG/Pt(2)/Ti($t_{Ti}$) and YIG/Pt(2)/Ge($t_{Ge}$) heterostructures, where $t_{Ti} = 0 - 30$ nm, and $t_{Ge} = 0 - 50$ nm. To study orbital current diffusion through Pt, we prepared YIG/Pt($t_{Pt}$)/X(10) structures, with X = Ti, or Au.

Single-crystal YIG films were grown onto 0.5 mm-thick (111)-oriented GGG (Gd$_3$Ga$_5$O$_{12}$) substrates by the Liquid Phase Epitaxy (LPE) technique in a three-zone vertical furnace. A Pt crucible containing the melt was placed in the central zone, with temperatures precisely controlled by three thermocouples connected to PID controllers. The films were grown at constant temperature, into a molten solution of PbO/B$_2$O$_3$ which is supersaturated with the garnet components Fe$_2$O$_3$ and Y$_2$O$_3$. Using a step-motor-driven mechanism, we lowered the GGG substrate into the furnace and held it just above the melt surface for thermalize before final immersion. During growth, the substrate was rotated at 50-150 rpm and, upon withdrawal, spun at 700 rpm to remove excess flux. The YIG films exhibited excellent crystalline quality, confirmed by their exceptionally narrow ferromagnetic resonance (FMR) linewidth ($\Delta H < 2.0$ Oe). For this study, rectangular samples with dimensions of 3.0 × 1.5 mm$^2$ and thickness ~400 nm were cut from the same single-crystal YIG film.

Metallic layers were deposited by DC sputtering at room temperature in a 3 mTorr argon atmosphere using the sputtering-up configuration. The substrate-target distance was fixed at 9 cm, with the plasma current fixed at 50 mA. Deposition rates were calibrated from film thicknesses determined X-ray reflectivity. The base pressure was 2.0×10$^{-7}$ Torr, and a five-minute pre-sputtering step was applied to remove oxides from the target surface.

For FMR and spin pumping-FMR (SP-FMR) measurements, samples were mounted on the tip of a PVC rod and inserted into a hole in the back wall of a rectangular microwave cavity operating in TE$_{102}$ mode at 9.41 GHz. The sample was positioned at the cavity wall, where the RF magnetic field is maximum and the RF electric field is minimum, thereby suppressing electric contributions from anisotropic magnetoresistance and spin rectification effects [55]. Field-scan absorption derivatives (dP/dH) were recorded using Helmholtz coils for field modulation at 1.2 kHz and lock-in detection. For DC voltage detection in the SP-FMR configuration, silver-painted electrodes were attached to the sample edges, and the voltage was measured with a nanovoltmeter at resonance. Field modulation was disabled during these measurements to avoid spurious induced signals.

### 2.1. Method for generating orbital current

The experimental techniques used to generate angular momentum currents can be broadly classified into direct and inverse methods. In the direct method, an electric current produces a transverse angular momentum current, whereas in the inverse method, an angular momentum current generates a transverse charge current. Pure spin currents are readily obtained through spin pumping, while the generation of pure orbital currents is more challenging. Using ISHE and IOHE, one can probe the interplay between spin and orbital currents in ferromagnet/normal metal (FM/NM) bilayers [42]. A seminal approach, developed in Ref. [42], employs a YIG/Pt bilayer. Spin pumping in YIG injects a pure spin current into an ultrathin Pt layer (2 nm), where strong SOC partially converts it into an orbital current, producing an intertwined spin–orbital current that can be transferred into a third material. The YIG/Pt system has thus become an effective orbital current injector for IOHE studies. While YIG is a common choice, other ferromagnetic materials such as Fe, Co, and Ni, or rare-earth doped garnets (e.g., Bi-YIG, Tm-YIG) can be used [35, 37-42, 56-58]. However, when using ferromagnetic metals instead of the insulating YIG, the interpretation of results becomes more complex. These metals themselves possess significant SOC (e.g., strong in Ni and Co compared to Fe [43,56,59]), meaning the spin pumping process can directly generate both spin and orbital currents within the FM itself. Furthermore, the presence of intrinsic galvanomagnetic effects (e.g., anomalous Hall effect, anisotropic magnetoresistance, etc.) in magnetic metals can obscure the signals from the injected currents, complicating a clear analysis of the experimental data.

Figure 3 illustrates the methodology. In (i), an insulating magnetic layer with very low damping, like YIG under FMR, generates an out-of-equilibrium spin polarization at its surface (red arrows). In (ii), the precessing magnetization $\vec{M}$ in YIG drives a spin current density at the YIG/Pt interface, containing both AC and DC components. The DC component, (red arrow, parallel to the applied field $\vec{H}$) injects a spin accumulation into Pt, where part of the spin current converts into a transverse charge current via the ISHE. The strong SOC in Pt also induces an orbital current that propagates through Pt and into an adjacent normal metal (iii), where orbital effects can be more directly probed. A similar process occurs when YIG spin dynamics are driven by a perpendicular thermal gradient through the spin Seebeck effect (SSE) [60], providing an alternative route for orbital current injection. Thus, YIG/Pt/NM heterostructures provide a versatile platform for generating and studying orbital currents, forming the basis for the experimental validation discussed below.

## 3. RESULTS AND DISCUSSIONS

### 3.1. SP-FMR signals for samples with CuO$_x$ interfaces

We present experimental results on orbital-to-charge conversion obtained from SP-FMR and SSE measurements. We first focus on interfacial orbital Rashba contributions at oxidized copper interfaces, then address bulk orbital transport in metals and semiconductors, and finally orbital diffusion in multilayers.

Motivated by the theoretical study of D. Go et al. [61], which predicted strong hybridization between Cu $p$ orbitals and O $d$ orbitals leading to a giant ORE, we demonstrate that orbital-to-charge conversion is strongly enhanced at oxidized Cu interfaces. To probe this effect, we investigated YIG/Pt(2) and YIG/Ti(4), with and without a CuO$_x$ capping layer. The capping layer was formed by depositing a 3 nm-thick island-like Cu film on the Pt and Ti, leaving the sample edges free for contacts, followed by natural oxidation in air, which drives oxygen diffusion to the Pt/Cu interface

Figure 4 (a) shows the SP-FMR signals for YIG/Pt(2)/CuO$_x$(3) (blue/red) and YIG/Pt(2) (green/orange), at $\phi = 0°$, $180°$ and $\phi = 90°$ (black). The inset defines $\phi$. Figure 4 (b) shows SSE signal for the same samples.

YIG/Pt(2)/CuO$_x$(3) sample (blue and red symbols for = 0°, $\phi = 180$, respectively), compared to YIG/Pt(2) (green and orange symbols). The CuO$_x$ layer enhances both SP-FMR and SSE signals by factors of ~ 4.5 and ~ 2.5, respectively, which is consistent with IORE at the Pt/CuO$_x$ interface [35]. Figures 4(c-d) compare YIG/Ti(4) and YIG/Ti(4)/CuO$_x$(3). In contrast to the Pt-based samples, the Ti-based heterostructures show: (i) unchanged SP-FMR amplitude after capping, (ii) signal magnitudes much smaller than in YIG/Pt(2)/CuO$_x$(3), and (iii) opposite polarity relative to Pt-based signals.

These results reveal both bulk and interfacial contributions to orbital current generation. In our orbital injection model, it is worth highlighting the contribution of two important interfaces: the YIG/Pt interface and the Pt/CuO$_x$ interface. The first interface provides important information about the transfer of spin angular momentum from YIG to Pt through spin pumping [62,63]. The spin current density $J_s(0)$ at the YIG/Pt interface is given by [64,65]:

$$J_s(0) = \frac{\hbar \omega g_{eff}^{\uparrow\downarrow}}{4\pi} \left(\frac{h_{RF}}{\Delta H}\right)^2 \frac{(H_r + 4\pi M_S)}{(2H_r + 4\pi M_S)^2} \frac{\omega}{\gamma} L(H - H_r), \qquad (1)$$

where $g_{eff}^{\uparrow\downarrow}$ is the real part of the effective spin mixing conductance and represents the efficiency of spin transfer at the interface, $\gamma$ is the gyromagnetic ratio, $\omega = 2\pi f$, and $h_{RF}$ are, respectively, the frequency and amplitude of the driving microwave magnetic field, $M_S$ is the saturation magnetization, and $L(H - H_r) = \Delta H^2 / [(H - H_r)^2 + \Delta H^2]$ is the normalized Lorentzian function being $H_r$ and $\Delta H$ the FMR field and linewidth, respectively. The spin accumulation at the YIG/Pt interface diffuses as a spin current throughout the Pt thickness [35,62,63]. A fraction of this spin current $\vec{J}_s$ is converted to a transverse charge current $\vec{J}_c$ due to ISHE [35,62,63], given by $\vec{J}_c = (2e/\hbar)\theta_{SH}(\vec{J}_s \times \hat{\sigma}_s)$ where $\theta_{SH}$ is the spin Hall angle, and $\hat{\sigma}_s$ is the spin polarization that is parallel to the applied magnetic field $\vec{H}$. Due to ISHE, a DC voltage arises along the length of the metal layer, measured between the contacts positioned at the two ends of the sample. Integration of the charge current density along $y$ and $z$ gives for the spin pumping current $I_{SP-FMR}$ [64,65]:

$$I_{SP-FMR} = w\lambda_s \left(\frac{2e}{\hbar}\right) \theta_{SH} \tanh(t_{NM}/2\lambda_s) J_s(0)\cos\phi, \qquad (2)$$

where $w$ is the width of the NM (Pt), $\lambda_s$ is the spin diffusion length, $\theta_{SH}$ is the spin Hall angle, which quantifies the efficiency of the spin-to-charge conversion process, $t_{NM}$ is the thickness of the NM layer, and $\phi$ is defined in the inset of Fig. 4(a).

Equation (2) describes the angular behavior of the results presented in Figure 4. Note that varying the angle from $\phi = 0°$ to $\phi = 180°$ inverts the polarity of the $I_{SP-FMR}$ signal in Pt and vanish at $\phi = 90°$. To extract typical values of $g_{eff}^{\uparrow\downarrow}$ and $J_s(0)$ from Equations (1) and (2) we considered the following parameters: $\gamma = 2.8$ GHz/kOe, $4\pi M_S = 1.76$ kG, $f = 9.41$ GHz, $\theta_{SH} = 0.01$ [66], $w = 1.5$ mm, $h_{RF} = 1$ Oe, and $H_r = 2.5$ kOe. Using these values, we obtain a spin mixing conductance of $g_{eff}^{\uparrow\downarrow} = 2 \times 10^{17}$ m$^{-2}$, and a spin diffusion length of $\lambda_s = 1.6$ nm. From these values we find $J_s(0) = 2.3 \times 10^7$ in units of $(e/\hbar)(A/m^2)$.

Samples with CuO$_x$(3) are not expected to exhibit large SP-FMR or SSE signals, since Cu has weak SOC [35,37,67] compared to Pt, leading to a negligible contribution when added to the YIG/Pt(2) bilayer. However, we observe a significant signal enhancement. Lorentzian fit shows a more than fivefold increase in signal magnitude, as shown in Fig. 4(a). This enhancement can only be attributed to the IORE in CuOx, as theoretically predicted by D. Go et al [61]. From these experimental results, we estimate the orbital Rashba parameter $\lambda_{IORE}$ for the Pt/CuOx interface, which quantifies the efficiency of the orbital-to-charge conversion mediated by Rashba-type states. The IOHE is defined by $\vec{J}_c = (2e/\hbar)\theta_{OH}(\vec{J}_L \times \hat{\sigma}_L)$, where $\hat{\sigma}_L$ is the orbital polarization. In materials with large SOC, $\hat{\sigma}_L$ can be inferred from the spin polarization $\hat{\sigma}_s$, since spin and orbital degrees of freedom are coupled. This equation mirrors the spin case, with spin variables replaced by their orbital counterparts. The parameter $\theta_{OH}$ is the orbital Hall angle, characterizing the orbital-to-charge conversion efficiency. In Pt, strong SOC generates an entangled spin current $\vec{J}_{LS}$ [35] that diffuses to the Pt(2)/CuOx interface, producing both an interfacial spin current $J_s^{int}$ and an interfacial orbital current $J_L^{int}$. The spin current is given by $J_s^{int} = \beta J_s(0)$, where in Pt $\beta \approx 0.23$, for $t_{Pt} = 2$ nm [41]. The orbital current can be approximated as $J_L^{int} \approx \eta_{SO} J_s^{int}$, where $-1 \leq \eta_{SO} \leq 1$ is the spin-to-orbital conversion factor. Using the value obtained for $J_s(0)$ and $\eta_{SO} \approx 1$ for Pt, we estimate $J_L^{int} \approx 5.3 \times 10^6 \ (e/\hbar)(A/m^2)$. From Figure 4(a), the difference between the SP-FMR peak signal of YIG/Pt(2)/CuOx(3) and YIG/Pt(2) is $I_{Peak}^{IORE} = 1236$ nA, corresponding to the charge current associated with IORE. The charge density is obtained as by $I_{Peak}^{IORE}/w$, yielding $J_c^{int} = 8.24 \times 10^{-4}$ A/m. The interfacial charge current density $\vec{J}_c^{int}$ due to IORE is given by [38]

$$\vec{J}_c^{int} = \lambda_{IORE}(\hat{z} \times \delta\vec{L}), \tag{3}$$

where the term $\delta\vec{L}$ represents the nonequilibrium orbital angular momentum density. Using the values obtained previously for the orbital current density $J_c^{int}$ at the Pt(2)/CuOx interface, we find $\lambda_{IORE} \approx J_c^{int}/J_L^{int} = 0.16$ nm. This value is in the order of those reported for Bi/Ag [26].

For the SSE signals (Figure 4 (b)), the spin density is generated from a thermal gradient $\vec{\nabla}T$, where the charge current due to ISHE obeys Equation (2), with $J_s(0)$ given by [60,64]

$$J_S(0) = -C_s \, \rho g_{eff}^{\uparrow\downarrow} \, \nabla T, \tag{4}$$

where $C_s$ is a coefficient that depends on the material parameters, temperature, and applied field intensity, and $\rho$ is a factor that represents the effect of the finite thickness of the FM layer [60,64]. Our SSE results are in agreement with Equation (4). The signal varies with the angle $\phi$, being positive for $\phi = 0°$ and inverting polarity for $\phi = 180°$. Furthermore, it presents the same dependence on the Pt layer thickness as the SP-FMR, saturating for sufficiently large $t_{Pt}$ and presenting a maximum for $t_{Pt} \approx 2$ nm. The SP-FMR and SSE presented similar results, both with a significant signal gain after the addition of the CuOx(3) layer. However, the difference in gains may be associated with the distinct microscopic characteristics of the spin currents generated at the YIG/Pt interface. In the SP-FMR, the spins precess in phase in the YIG, resulting in a coherent injection of spin current into the Pt, while in the SSE, the spins exhibit a phase shift in the precession due to the applied thermal gradient. This mechanism is not yet fully understood and requires further investigation.

Interfacial orbital dynamics are no longer observed in the YIG/Ti/CuOx(3) samples, evidencing the absence of orbital contribution in these systems. This occurs because Ti, due to its extremely weak SOC [38,67], is not capable of generating significant orbital current and, in this case, only spin current is transported to the Ti/CuO$_x$ interface. Figures 4 (c-d) illustrate this difference. The observed SP-FMR signal obeys Equation (4) and is attributed only to the weak ISHE in the Ti layer.

To investigate the interplay between spin and orbital angular momentum in all-metal structures, we performed SP-FMR experiments on multilayers designed with different stacking order to allow opposite spin current injection directions. Figure 5(a) shows the SP-FMR signals obtained for SiO$_2$/Py(5)/Pt(4) (blue symbols) and SiO$_2$/Pt(4)/Py(5) (red symbols). The former exhibits a positive signal, while the latter exhibits a negative signal, consistent with the inversion of the Py deposition order relative to Pt. These results agree with the ISHE relation $\vec{J}_c \propto (\vec{J}_s \times \hat{\sigma}_s)$, where reversing $\vec{J}_s$ changes the signal polarity. The theoretical fit (solid line) combines symmetric and antisymmetric components associated with the metallic Py layer, with dashed lines indicating each contribution separately. The inset shows the Py(5) self-conversion signal under the same experimental conditions ($\phi = 0°$, RF power of 110 mW), revealing a negative symmetric component. This contribution explains the larger signal observed in the SiO$_2$/Pt(4)/Py(5) sample compared to SiO$_2$/ Py(5)/Pt(4). Thus, the total signal results from the superposition of two contributions: ISHE in Pt(4) and the self-conversion in Py(5). When Py is on top, the negative Pt ISHE adds to the Py self-conversion signal, whereas when Py is at the bottom, the positive Pt ISHE partially cancels it.

Figure 5(b) shows the SP-FMR signal for the SiO$_2$/Py(5)/Pt(4)/CuO$_x$(3) sample, revealing a positive symmetric component. This is attributed to the combined effects of ISHE in Pt and IORE at the Pt/CuO$_x$ interface. The symmetric component in this sample is approximately three times larger than that of the reference SiO$_2$/Py/Pt sample in Figure 5(a). On the other hand, the sample with inverted stacking order, SiO$_2$/CuO$_x$(3)/Pt(4)/Py(5) in Figure 5(c), shows a symmetric component 1.6 times smaller than the reference sample. After theoretical fitting to isolate the symmetric contributions, we subtracted the signals of the samples with and without the CuO$_x$ layer. The results, shown in Figures 5(d) and 5(e), yield a positive difference of 19 nA when CuO$_x$ is below Pt and 20 nA when it is above. This indicates that the stacking order does not alter the polarity of the IORE-generated charge current, confirming that the dominant mechanisms are the ISHE of Pt and the IORE of CuO$_x$. A bar graph comparing these differences is presented in Figure 5(f). We therefore conclude that the IORE in our samples arises directly from the *p-d* hybridization in the Cu/O orbitals, being independent of the relative position of Pt. In this context, Pt acts only as an inert layer, responsible for the orbital current injection (either from top to bottom or from bottom to top). This behavior is fully consistent with Equation (3), which depends exclusively on the OAM density at the interface, not on the direction of the orbital current.

To date, our study has focused on orbital current generation through the SOC in Pt, generally avoiding ferromagnetic metals to minimize galvanomagnetic contributions. However, the same SP-FMR setup can detect electrical signals in FM monolayers, where the intrinsic SOC of the FM can induce orbital accumulation at an interface, convertible into an electrical signal. To investigate this effect and explore properties like the *LS* coupling strength, we fabricated two series of samples: (i) SiO$_2$/FM(10) with FM = Py, Co, Fe, Ni; and (ii) the same structures with an added CuO$_x$(3) layer. FMR measurements showed nearly identical linewidths with and without CuO$_x$. Figures 6(a-d) show the FMR absorption curves at 110 mW RF power, while Figures 6(e-h) show the self-conversion signals for both sample types. Significant changes were observed only for SiO$_2$/Co(10)/CuO$_x$(3) and SiO$_2$/Ni(10)/CuO$_x$(3). We attribute this to the IORE: the moderate SOC in Co and Ni favors the OAM accumulation at the FM/CuO$_x$ interface, generating a positive orbital-to-charge conversion signal of opposite polarity to the FM autoconversion signal. In contrast, The Py and Fe samples showed no significant changes, consistent with their weak SOC [59]. Figure 6(i) summarizes the peak values of the symmetric SP-FMR component for all samples, demonstrating that OAM can be injected using FM metals with moderate SOC. This paves the way for studying the inverse orbital Hall effect (IOHE) in FM/NM structures. The insets of Figure 6(i) compares the SP-FMR signal of SiO$_2$/Co(10)/Ti(20) with that of YIG/Ti(4), both measured at an RF power of 110 mW. A symmetric IOHE component of 15 nA was found for Co/Ti, whereas the ISHE signal in YIG/Ti was only ~0.5 nA, which is nearly two orders of magnitude smaller. This confirms a strong IOHE in Ti, activated specifically by orbital current injection.

### 3.2. SP-FMR and SSE signals in metals and semiconductors

In this section, we present the results of the SP-FMR and SSE signals for YIG/Pt(2)/Ti($t_{Ti}$) and YIG/Pt(2)/Ge($t_{Ge}$) films. Figure 7(a) shows the variation of the peak signal ($I_{Peak}$) obtained by the SP-FMR technique for the Ti sample (red symbols). The solid curve is the theoretical fit to the experimental data, as discussed below. We clearly observe an increase in the SP-FMR signal when increasing the Ti thickness until its saturation. The inset shows two line shapes for $t_{Ti} = 0$ nm and $t_{Ti} = 30$ nm, for $\phi = 0°$ and RF power at 15 mW. Figure 7(b) shows the SSE signals as a function of Ti layer thickness, which exhibited the same behavior as SP-FMR. Each data point is obtained from the difference between the saturation regions. The inset shows the lineshapes obtained for $\phi = 0°$ and a thermal gradient fixed at 10 K, which shows a significant increase in the signal.

According to Ref. [50], Ti has $\sigma_{OH} = 4304 \left(\frac{\hbar}{e}\right)(\Omega \cdot cm)^{-1}$ and $\sigma_{SH} = -17 \left(\frac{\hbar}{e}\right)(\Omega \cdot cm)^{-1}$. These values suggest that orbital effects should be more pronounced than spin effects in Ti. In Figure 7, several remarkable features emerge: (i) the orbital current injected into Ti is converted into a transverse charge current via IOHE, obeying the equation $\vec{J}_c \propto (\vec{J}_L \times \hat{\sigma}_L)$. (ii) The resulting SP-FMR and SSE signal exhibits significant gains compared to that of YIG/Pt(2), as shown by the curves in Figures 7(a) and (b), which cannot be explained by the spin analogue due to the negligible ISHE of Ti, i.e., $\theta_{SH}^{Ti} \ll \theta_{OH}^{Ti}$.

Similarly, Ge also has significant orbital conductivity. A recent theoretical study [68] predicts that pure Ge exhibits an orbital Hall conductivity $\sigma_{OH} = -\left(\frac{\hbar}{e}\right) 1270 \ (\Omega \cdot cm)^{-1}$ and a spin Hall conductivity $\sigma_{SH} = \left(\frac{\hbar}{e}\right) 1.6 \times 10^{-1} (\Omega \cdot cm)^{-1}$, making it an excellent candidate for investigating orbital effects. Materials with a negative orbital Hall angle ($\theta_{OH} < 0$), such as Ge, are particularly valuable for disentangling orbital and spin contributions in hybrid systems. To explore the negative IOHE in Ge, we employed the same methodology. Samples with the structure YIG/Pt(2)/Ge($t_{Ge}$) were fabricated, and both SP-FMR, for an RF power of 43 mW (Figure 8 (a)), and SSE, for a thermal gradient of 10 K (Figure 8 (b)), signals were analyzed. Given that Ge exhibits $\theta_{SH} \ll 1$ and $\theta_{SH} > 0$ [39], the ISHE contribution from Ge can be neglected. Figure 8 (a-b) shows a clear reduction in the signal upon adding the Ge layer on top of YIG/Pt(2), which is attributed to the negative orbital conductivity of Ge. This orbital contribution opposes the ISHE generated by the Pt layer, thereby reducing the net signal. For $t_{Ge} = 50$ nm, the signal becomes nearly zero, requiring a 100-fold scale amplification to be visible on the same scale as YIG/Pt(2).

To obtain important physical parameters, such as the polarity of the orbital Hall angle $\theta_{OH}$, its magnitude, and the orbital diffusion length $\lambda_L$, we used a model [37,69] based on the diffusion of spin angular momentum (SAM) and OAM in a FMI/NM1/NM2 heterostructure, where FMI represents a magnetic insulator, in our case YIG, NM1 a material with strong SOC, for which we used Pt, and NM2 is a material with a dominant orbital response over the spin contribution, such as Ti, and Ge. Specifically, the model considers the generation and diffusion of spin and angular momenta and their interconversion mediated by SOC. In the YIG/NM1/NM2 three-layer, the magnetization precession in YIG injects a spin current into NM1, where the diffusion of spins and orbital chemical potentials $\mu_{1S}$ and $\mu_{1L}$ is governed by

$$\frac{d^2\mu_{1S}}{dy^2} = \frac{\mu_{1S}}{\lambda_{1S}^2} - \frac{\mu_{1L}}{\lambda_{1LS}^2}, \tag{5a}$$

$$\frac{d^2\mu_{1L}}{dy^2} = \frac{\mu_{1L}}{\lambda_{1L}^2} - \frac{\mu_{1S}}{\lambda_{1LS}^2}, \tag{5b}$$

where $\lambda_{1S(L)}$ is the spin (orbital) diffusion length, and $\lambda_{1LS}$ is a phenomenological term [37,69] representing the conversion between spins and orbital inside NM1.

In NM2 where there is no interconversion between spin and orbital currents, the diffusion equations for the spin and orbital chemical potentials $\mu_{2S}$ and $\mu_{2L}$ are

$$\frac{d^2\mu_{2S}}{dy^2} = \frac{\mu_{2S}}{\lambda_{2S}^2}, \tag{6a}$$

$$\frac{d^2\mu_{2L}}{dy^2} = \frac{\mu_{2L}}{\lambda_{2L}^2}. \tag{6b}$$

The boundary conditions for the diffusion problem are given by the pumped spin current at the FMI/NM1 interface ($y=0$)

$$J_{1S}(0) = -\frac{\sigma_1 \hbar}{2e^2} \partial_y \mu_{1S}\Big|_{y=0}, \tag{7a}$$

and the absence of a pumped orbital current

$$\partial_y \mu_{1L}\Big|_{y=0} = 0. \tag{7b}$$

The other boundary conditions are determined by the continuity of the spin and orbital chemical potentials and the corresponding spin and orbital current at the NM1/NM2 interface, and the vanishing of the currents at the NM2 outer boundary. The solutions of the coupled Equations (5a) and (5b) are

$$\mu_{1S}(y) = A_S \cosh(y/\lambda_1) + B_S \sinh(y/\lambda_1) + C_S \cosh(y/\lambda_2) + D_S \sinh(y/\lambda_2), \tag{8a}$$

$$\mu_{1L}(y) = A_L \cosh(y/\lambda_1) + B_L \sinh(y/\lambda_1) + C_L \cosh(y/\lambda_2) + D_L \sinh(y/\lambda_2), \tag{8b}$$

where,

$$\frac{1}{\lambda_{12}^2} = \frac{1}{2}\left[\left(\frac{1}{\lambda_{1S}^2} + \frac{1}{\lambda_{1L}^2}\right) \pm \sqrt{\left(\frac{1}{\lambda_{1S}^2} - \frac{1}{\lambda_{1L}^2}\right)^2 + 4\frac{1}{\lambda_{1LS}^4}}\right], \tag{9}$$

are the combined spin-orbital diffusion lengths resulting from the coupling of the spin and orbital degrees of freedom introduced by $\lambda_{1LS}$ [69]. Imposing the boundary conditions, the spin and orbital currents in NM1 and NM2, respectively (see supplementary material) are

$$\begin{aligned}J_{1S}(y) &= \left[-\frac{\mathbb{A}_S}{\lambda_1}\sinh\left(\frac{y}{\lambda_1}\right) - \frac{\mathbb{C}_S}{\lambda_2}\sinh\left(\frac{y}{\lambda_2}\right) + \frac{1}{(1-\gamma_1/\gamma_2)}\cosh\left(\frac{y}{\lambda_1}\right)\right.\\ &\left.+ \frac{1}{(1-\gamma_2/\gamma_1)}\cosh\left(\frac{y}{\lambda_2}\right)\right]J_{1S}(0).\end{aligned} \tag{10}$$

$$J_{2L}(y) = \left(\frac{\sigma_2}{\sigma_1}\right)J_{1S}(0)\frac{\mathbb{A}_{2L}}{\lambda_{2L}}\sinh[(d_1 + d_2 - y)/\lambda_{2L}], \tag{11}$$

where $\mathbb{A}_S$, $\mathbb{C}_S$ and $\mathbb{A}_{2L}$ are functions of the parameters defined above (supplementary material).

By the assumptions made above, we consider that the full load current measured in SP-FMR is a composition of only the ISHE in the Pt layer and the IOHE in the NM2 layer. In this case, the charge current densities in Pt $\langle j_c^{Pt}\rangle$, and NM2 $\langle j_c^{NM2}\rangle$ are obtained by integrating Eqs. (10) and (11) (see supplementary material):

$$\langle j_c^{Pt}\rangle = \left(\frac{2e}{\hbar}\right)\frac{\theta_{SH}}{d_1}\int_0^{d_1} J_{1S}(y)\, dy, \tag{12}$$

and

$$\langle j_c^{NM2}\rangle = \left(\frac{2e}{\hbar}\right)\frac{\theta_{OH}}{d_2}\int_{d_1}^{d_1+d_2} J_{2L}(y)\, dy. \tag{13}$$

From the theoretical fits (Figures 7 and 8) we obtain: $\lambda_L^{Ti} = 3.5$ nm, $\lambda_L^{Ge} = 3.8$ nm, $\theta_{OH}^{Ti} = 0.1$, $\theta_{OH}^{Ge} = -0.029$, $\lambda_s^{Pt} = 1.6$ nm, $\theta_{SH}^{Pt} = 0.01$ (for more details, see Table 1 and the supplementary material). The fitted parameters indicate a clear predominance of orbital over spin contributions in the investigated materials. The extracted orbital diffusion lengths, and orbital Hall angles for Ti and Ge, imply efficient orbital transport and conversion in both systems. These magnitudes substantially exceed the spin Hall angle of our Pt injector, consistent with theoretical predictions and recent experimental reports that point to large intrinsic orbital Hall conductivities in light metals and semiconductors [50,68]. The opposite sign of $\theta_{OH}$ for Ge provides a convenient experimental handle to disentangle orbital and spin contributions: the positive $\theta_{OH}$ of Ti adds constructively to the ISHE signal of Pt, while the negative $\theta_{OH}$ of Ge partially cancels it, explaining the thickness-dependent trends observed in Figures 7-8.

### 3.3. SP-FMR signals in Pt($t_{Pt}$)/NM

To directly probe the orbital diffusion length and its dependence on the Pt interlayer thickness, we fabricated a series of heterostructures YIG/Pt($t_{Pt}$)/NM (NM=Ti(10) or Au(10)) in which the orbital currents are generated via spin pumping and propagate through Pt interlayer of varying thickness (0-10 nm). These trilayers provide access to orbital contributions through IOHE, which is not possible using only YIG/Pt bilayers where the signal is dominated by the ISHE.

Figure 9(a) compares the reference YIG/Pt($t_{Pt}$) samples (gray symbols) with YIG/Pt($t_{Pt}$)/Ti(10) (red symbols). The Ti-capped structures show a strong enhancement at small $t_{Pt}$, followed by saturation with increasing $t_{Pt}$, consistent with the spin-orbital diffusion limit. For $t_{Pt} > 4$ nm, the ISHE contribution is already saturated. Figure 9 (b) shows the signals for YIG/Pt($t_{Pt}$)/Au(10) (red symbols). It is observed that, for sufficiently large Pt thicknesses, both signals (Figure 9 (a-b)) reach saturation. However, in the Ti sample, saturation occurs above the reference value, while in the Au sample it occurs below. This contrasting behavior can be explained by the orbital Hall angles of the capping layers: positive for Ti ($\theta_{OH} > 0$) and negative for Au ($\theta_{OH} < 0$) [50]. In the Ti samples, the IOHE-generated current adds constructively to the Pt ISHE, yielding enhanced signals, whereas in the Au samples, the IOHE-generated current has the opposite polarity and partially cancels the Pt ISHE contribution. Importantly, this reduction cannot be attributed to spin effects, since Au has a positive spin Hall angle that would increase the signal rather than decrease. Theoretical fits using the model described in supplementary material yield $\theta_{SH}^{Ti} \sim 0$, and $\theta_{OH}^{Ti} = 0.1$, $\theta_{SH}^{Au} = 0.0004$, and $\theta_{OH}^{Au} = -0.003$. These values demonstrate that the experimental trends in YIG/Pt/Ti and YIG/Pt/Au cannot be explained by spin effects alone, and provide clear evidence of the role of orbital currents.

## 5. CONCLUSIONS

We have conducted a systematic investigation of orbital-to-charge conversion in metallic and semiconducting heterostructures using spin pumping and spin Seebeck techniques. Our experiments demonstrate that orbital contributions not only coexist with spin effects but can dominate them in several systems. In particular, interfacial orbital hybridization at naturally oxidized Cu layers strongly enhances the conversion efficiency, while the bulk Hall orbital responses in Ti, Ge, and Au provides direct access to orbital diffusion lengths and the polarity of their orbital Hall angles. The extracted values confirm that orbital contributions in these systems can exceed those of spin, even when mediated by Pt. A key outcome of this work is the observation of orbital signals with positive and negative polarity, which enables a clear disentanglement of orbital and spin channels in hybrid structures. Taken together, our results establish direct experimental evidence for the inverse orbital Hall and inverse orbital Rashba effects, highlighting their fundamental role in orbital transport and conversion. These findings open promising directions for the design of orbitronic devices that exploit orbital angular momentum as an efficient and versatile information carrier.

**ACKNOWLEDGMENTS**

This research is supported by Conselho Nacional de Desenvolvimento Científico e Tecnológico (CNPq), Coordenação de Aperfeiçoamento de Pessoal de Nível Superior (CAPES) (Grant No. 1575/2024), Financiadora de Estudos e Projetos (FINEP), Fundação de Amparo à Ciência e Tecnologia do Estado de Pernambuco (FACEPE), Universidade Federal de Pernambuco, Multiuser Laboratory Facilities of DFUFPE, Fundação de Amparo à Pesquisa do Estado de Minas Gerais (FAPEMIG) - Rede de Pesquisa em Materiais 2D and Rede de Nanomagnetismo, and INCT of Spintronics and Advanced Magnetic Nanostructures (INCT-SpinNanoMag), CNPq Grant No. 406836/2022-1, and in Chile by Fondo Nacional de Desarrollo Científico y Tecnológico (FONDECYT) Grant No. 1130705, and FONDEQUIP projects EQM180103, EQM190136 and EQM210105.

## Inverse Orbital Rashba Effect (IORE)

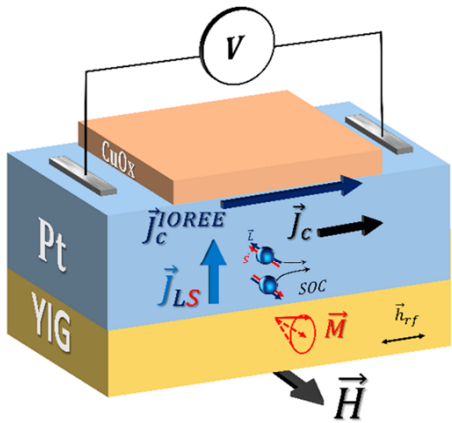

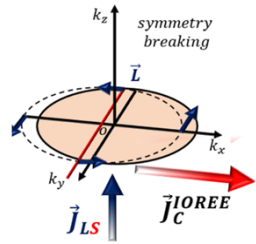

## Orbital Rashba Effect (ORE)

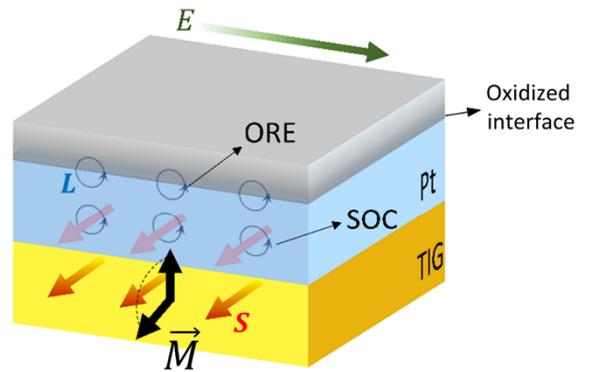

## Inverse Orbital Hall Effect (IOHE)

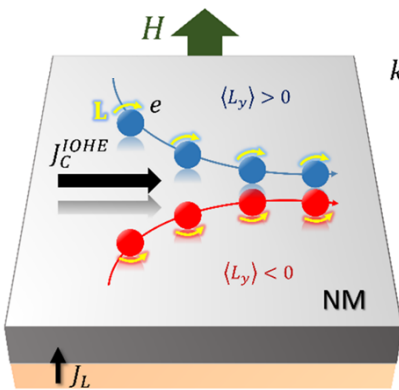

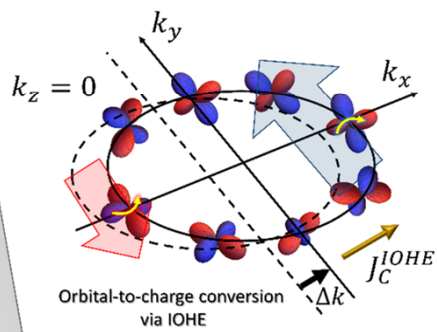

## Orbital Hall Effect (OHE)

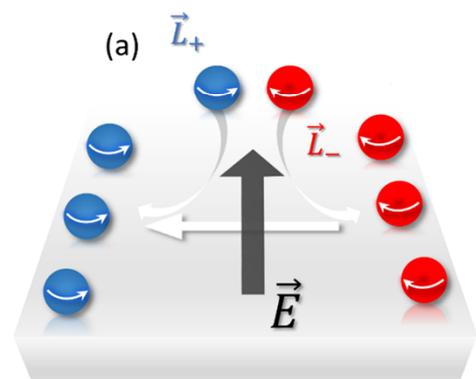

**Figure 1. Method for orbital current injection.** Direct orbital effects (ORE, OHE) and Inverse orbital effects (IORE, IOHE).

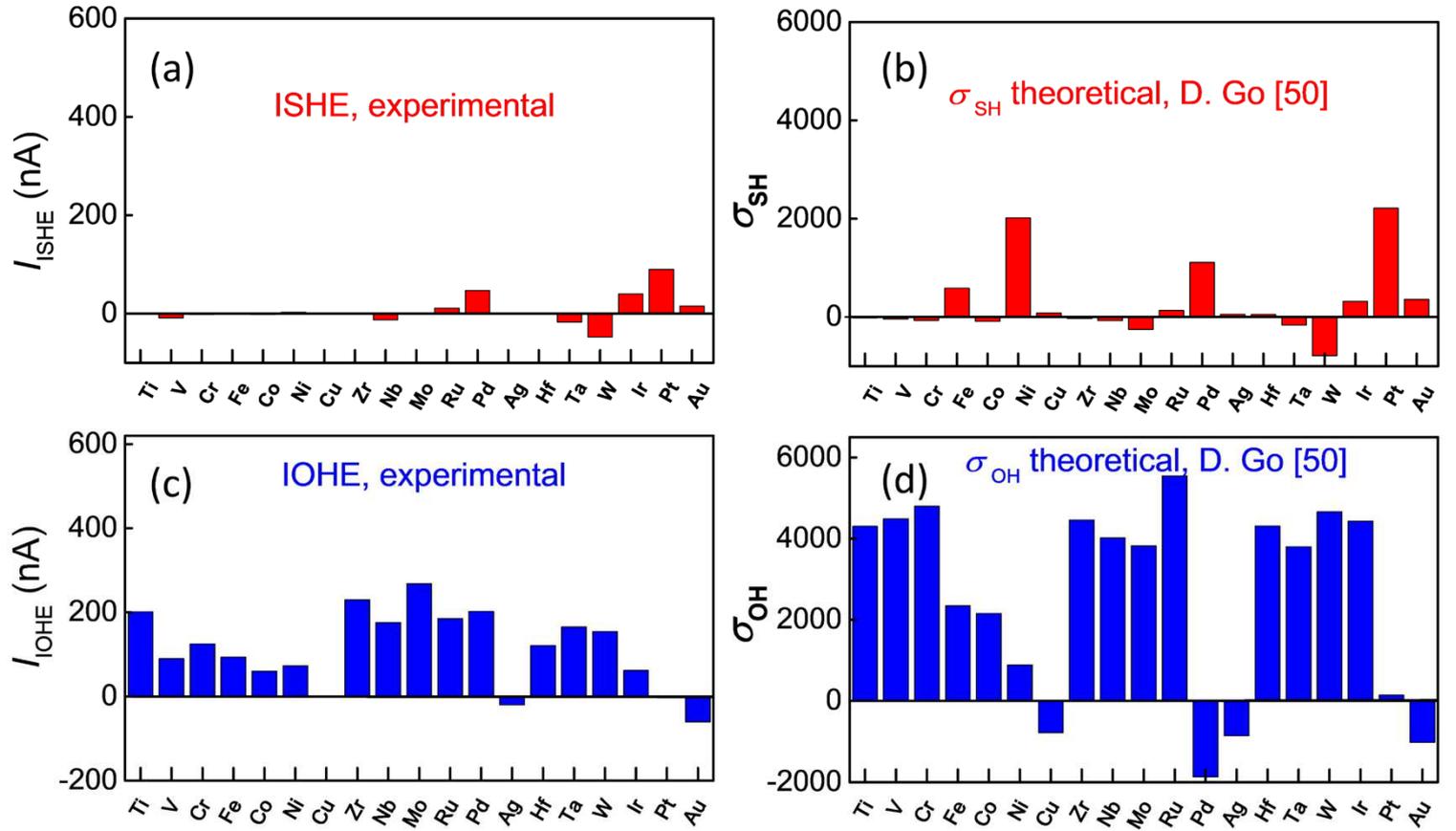

**Figure 2. Comparison between experimental signals and theoretical predictions for transition metals.** (a) Experimental ISHE current $I_{ISHE}$ [41], proportional to the spin Hall conductivity ($\sigma_{SH}$). (b) Theoretical spin Hall conductivity ($\sigma_{SH}$) from D. Go [50]. (c) Experimental IOHE current ($I_{IOHE}$) [41], proportional to the orbital Hall conductivity ($\sigma_{OH}$). (d) Theoretical orbital Hall conductivity ($\sigma_{OH}$) from D. Go [50].

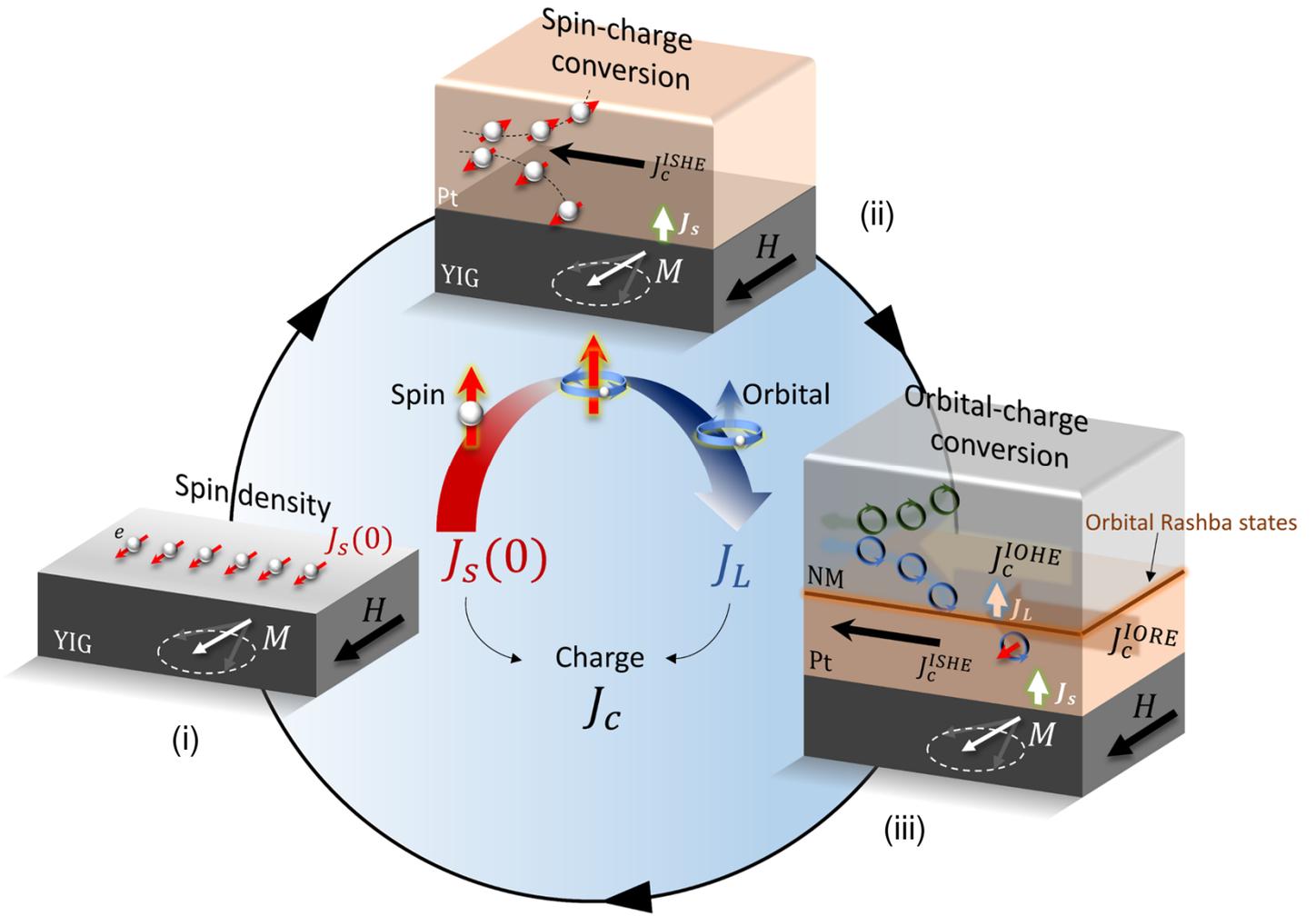

**Figure 3. Generation of orbital currents via spin–orbit coupling.** (i) The precessing magnetization $\vec{M}$ in YIG film, driven by FMR or a thermal gradient, creates a nonequilibrium spin accumulation at the interface ($\vec{J}_s(0)$). (ii) This spin accumulation drives a spin current $\vec{J}_s$ across the YIG/Pt interface, where part of it converts into a transverse charge current through the ISHE. The strong SOC in Pt also induces an orbital current ($\vec{J}_L$) that propagates through the metallic layer. (iii) In YIG/Pt/NM heterostructure, this orbital current reaches the NM layer and is converted into a measurable charge current ($\vec{J}_c$) via either the interfacial IORE or bulk IOHE.

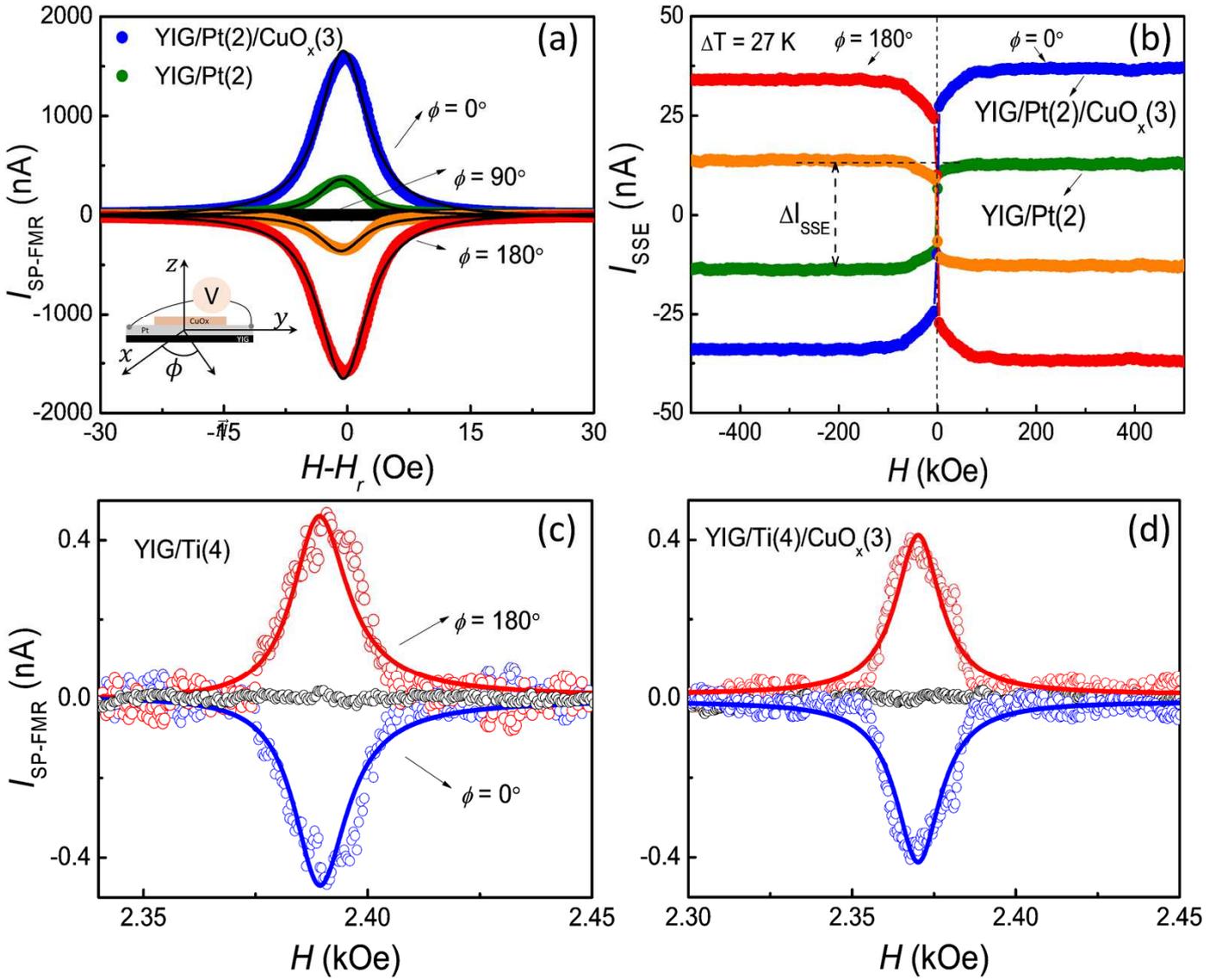

**Figure 4. SP-FMR and SSE signals of YIG-based heterostructures.** (a) SP-FMR signals for YIG/Pt(2) (green and orange symbols) and YIG/Pt(2)/CuO$_x$(3) (blue and red symbols), measured at $P_{RF} = 110$ mW. The inset defines the angle ϕ. (b) SSE signal for the same samples under a thermal gradient of $\Delta T = 27$ K, for $\phi = 0°, 180°$. (c-d) SP-FMR signals for YIG/Ti(4) and YIG/Ti(4)/CuO$_x$(3), respectively, measured for $\phi = 0°$ (blue symbols) and $\phi = 180°$ (red symbols). The solid lines represent symmetric Lorentzian fits to the data.

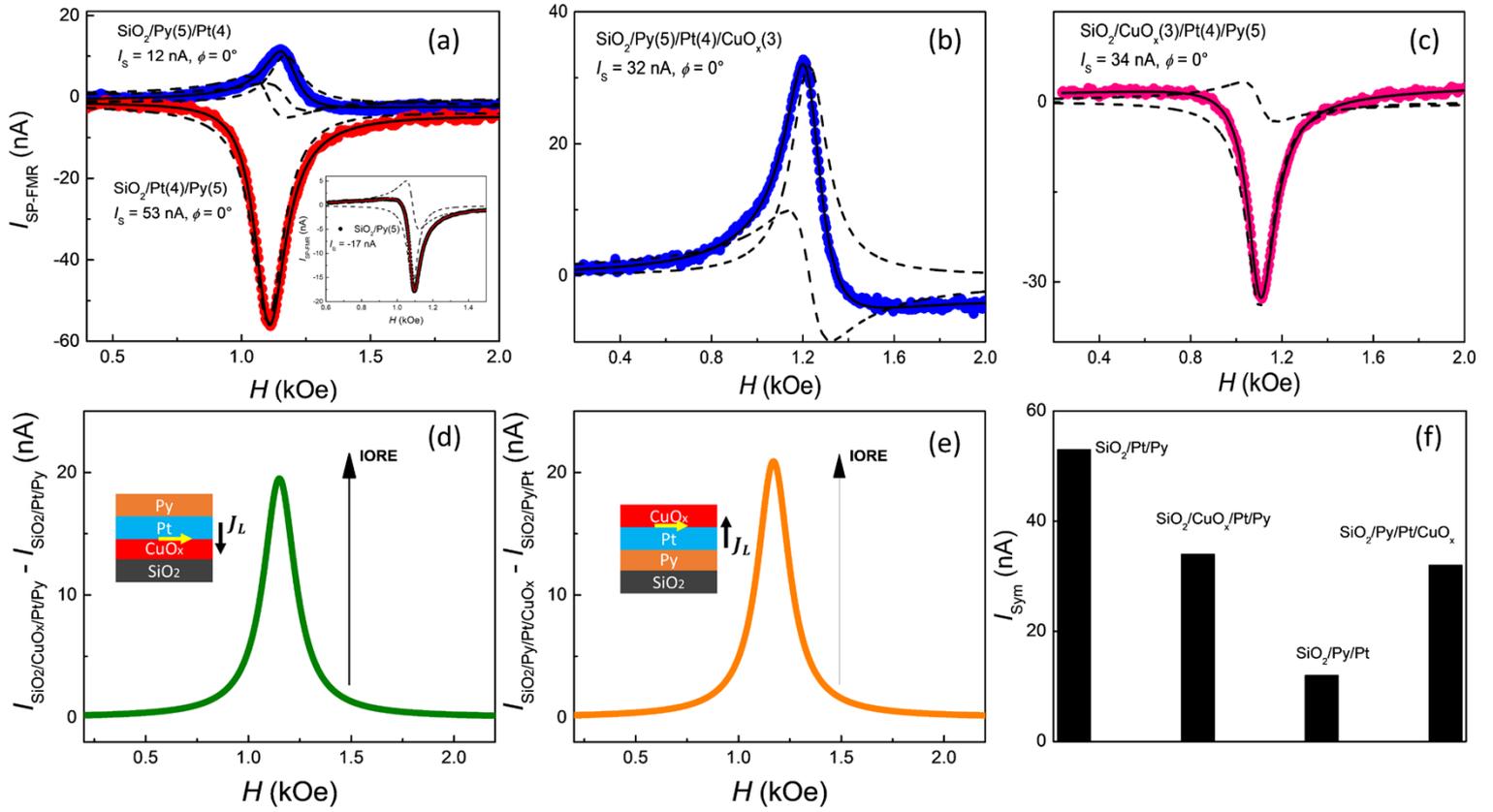

**Figure 5. SP-FMR signals in Pt/Py-based stacks with and without a CuO$_x$(3) layer.** (a) Reference signals for SiO$_2$/Pt(4)/Py(5) (red) and SiO$_2$/Py(5)/Pt(4) (blue), with a dashed green curve showing the symmetric and antisymmetric fit. (b) Signal for the stack with CuO$_x$ above Pt: SiO$_2$/Py(5)/Pt(4)/CuO$_x$(3). (c) Signal for the stack with CuO$_x$ below Pt: SiO$_2$/CuO$_x$(3)/Pt(4)/Py(5). (d-e) Curves obtained from the subtraction of the reference signals: $I_{CuOx(3)/Pt(4)/Py(5)} - I_{Pt(4)/Py(5)}$; and $I_{Py(5)/Pt(4)/CuOx(3)} - I_{Pt(4)/Py(5)}$. (f) Comparison of the magnitudes of these signals, highlighting enhancement or suppression depending on the CuO$_x$(3) configuration. All measurements were performed at $\phi = 0°$ with a fixed RF power of 110 mW. The spin current polarization direction was kept constant, with only the injection direction inverted.

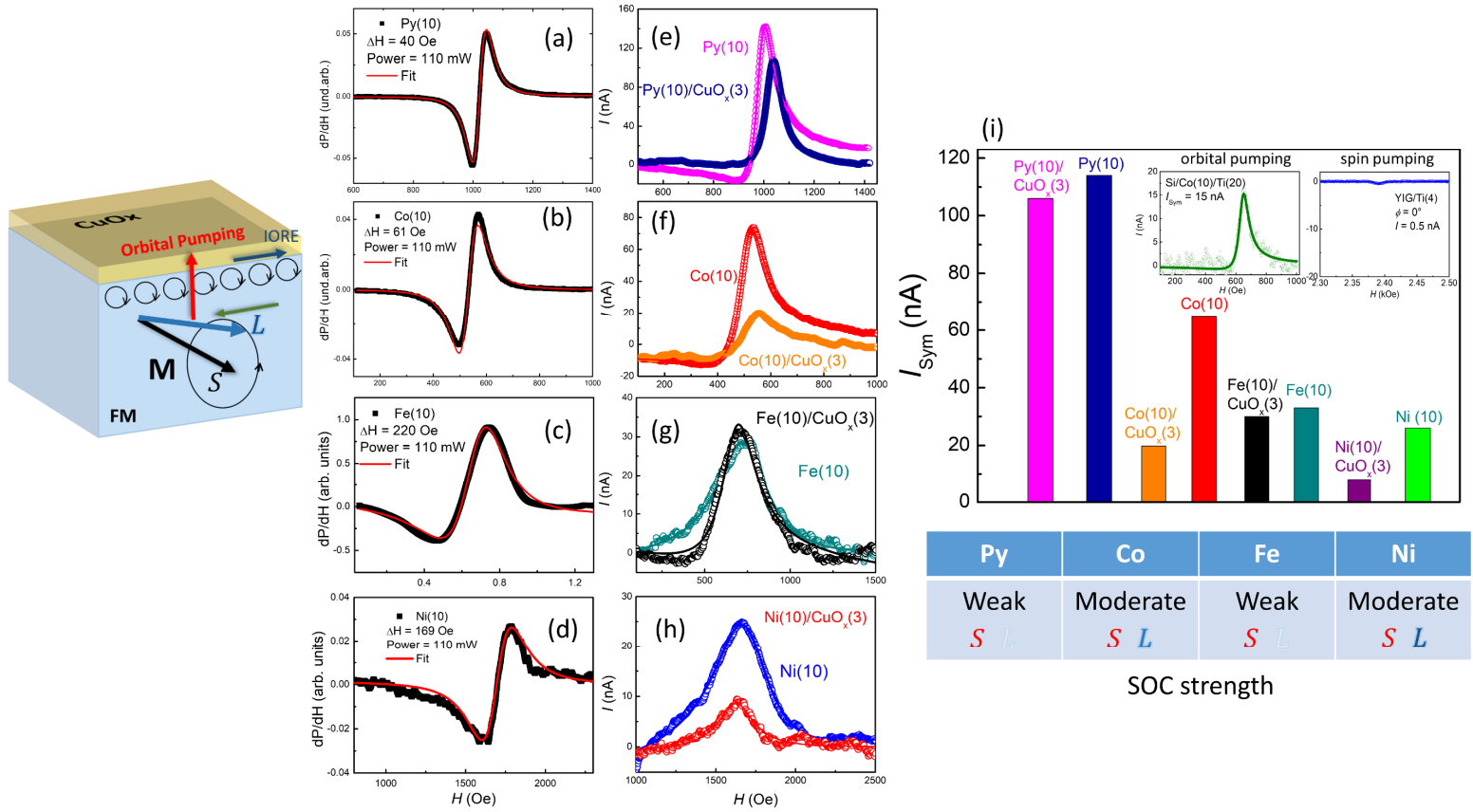

**Figure 6. Influence of CuO$_x$ on self-conversion signals in magnetic metals.** (a-d) FMR absorption curves in SiO$_2$/Py(10), SiO$_2$/Co(10), SiO$_2$/Fe(10), and SiO$_2$/Ni(10), respectively. The fit of the FMR absorption curve for Fe was performed with a Dyssonian function, suitable due to the more pronounced dispersive contribution associated with electrical conductivity, which generates asymmetry in the FMR spectrum. (e-h) Signals from the samples with the CuO$_x$(3) capping layer. (i) Comparison of the spin rectification signals and signals associated with the presence of the CuO$_x$(3) layer. The inset shows the orbital and spin pumping using SiO$_2$/Co(10)/Ti(20) and YIG/Ti(4), respectively. The table elucidates the SOC intensity in the materials used.

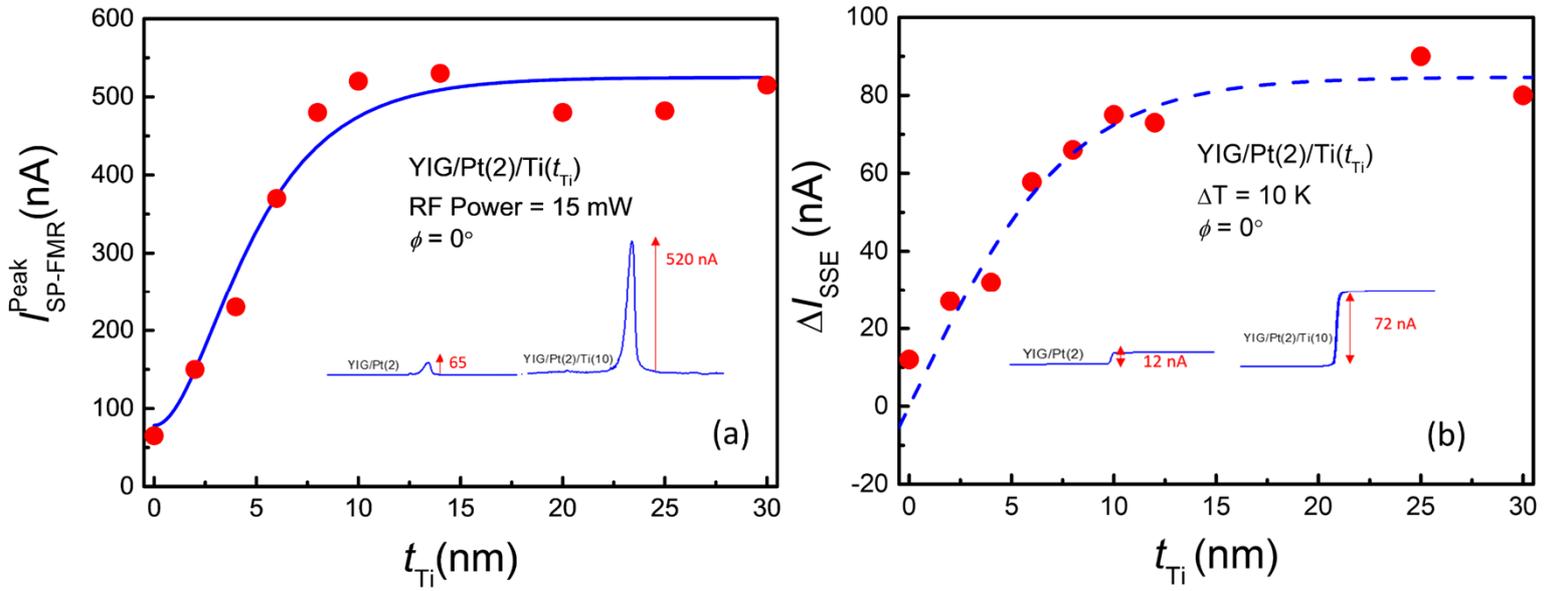

**Figure 7. SP-FMR and SSE in YIG/Pt(2)/Ti($t_{Ti}$) heterostructures.** (a) SP-FMR peak currents (red symbols) as a function of the Ti layer thickness (0 – 30 nm), measured at $\phi = 0°$ with an RF power of 15 mW. All samples exhibit the same FMR resonance field (~2.5 kOe at 9.41 GHz). The solid blue line represents the theoretical fit, and the inset shows representative SP-FMR spectra. (b) SSE signals (red symbols) measured in the same series under a fixed thermal gradient of $\Delta T = 10\ K$. $\Delta I_{SSE}$ was extracted from the difference between the upper and lower saturation levels. The dashed blue line is a visual guide, and inset displays representative SSE signals.

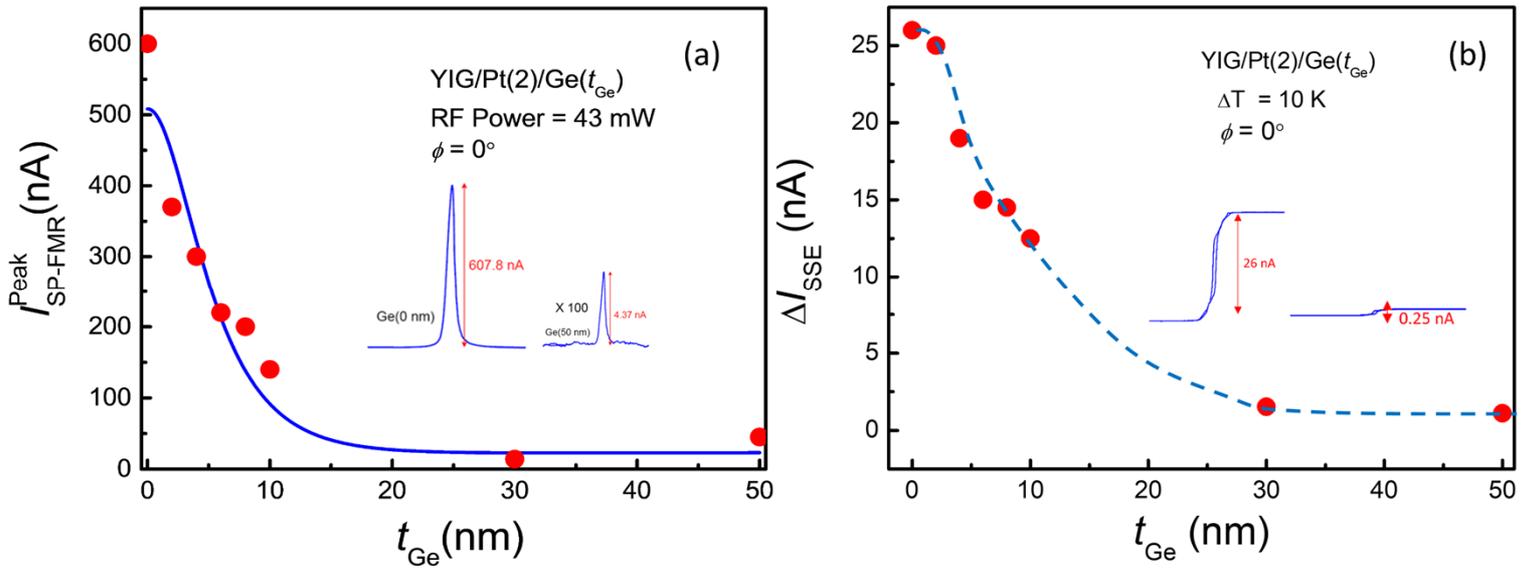

**Figure 8. SP-FMR and SSE signals in YIG/Pt(2)/Ge($t_{Ge}$) heterostructures.** (a) SP-FMR peak currents (red symbols) as a function of the Ge (0 – 50 nm), measured at $\phi = 0°$ with an RF power of 43 mW. The solid blue line represents the theoretical fit. Insets show representative SP-FMR spectra for Ge(0 nm) and Ge(50 nm). (b) SSE signals measured (red symbols) measured in the same sample series under a fixed thermal gradient of $= 10\ K$ at $\phi = 0°$. The dashed blue curve is a visual guide. The insets show representative SSE curves.

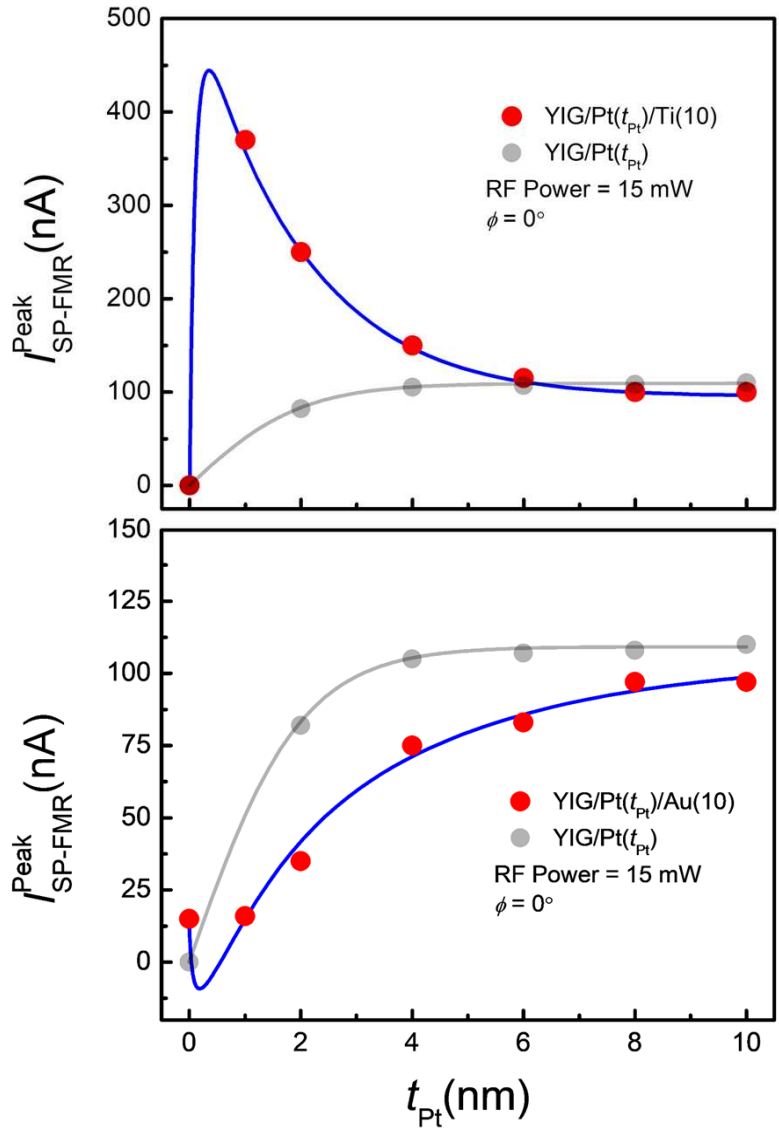

**Figure 9. SP-FMR signals as a function of Pt thickness.** (a) SP-FMR signals for YIG/Pt($t_{Pt}$)/Ti(10) (red symbols), compared with the reference YIG/Pt($t_{Pt}$) (gray symbols). The solid blue curve represents the theoretical fit. The Ti-capped samples exhibit a strong enhancement at small $t_{Pt}$, followed saturation above 4 nm. (b) SP-FMR signals for YIG/Pt($t_{Pt}$)/Au(10) (red symbols), also compared with YIG/Pt($t_{Pt}$) (gray symbols). In this case, the signal saturates below the reference values. For both systems, saturation occurs for sufficiently large Pt layer thicknesses, but with opposite trends depending on the sign of the orbital Hall angle of the capping layer.

**Table 1. Physical parameters obtained from the developed theoretical model for IOHE.** The table summarizes the spin diffusion length ($\lambda_S$), orbital diffusion length ($\lambda_L$), spin Hall angle ($\theta_{SH}$), and orbital Hall angle ($\theta_{SH}$) for Pt, Ti, Ge, and Au. The spin–orbital coupling length $\lambda_{LS}$, which governs spin–orbital interconversion, lies between 1.4 nm and 1.7 nm. For Au, the orbital diffusion length was not extracted since the Au layer thickness was fixed at 10 nm, whereas Ti and Ge were studied as a function of thickness. The spin Hall angle of Ge and Ti were approximated to zero, consistent with our experimental results. The effective spin mixing conductance ($g_{eff}^{\uparrow\downarrow}$) ranged from $2 \times 10^{16}$ m$^{-2}$ and $8 \times 10^{16}$ m$^{-2}$ across different sample series. The variations in extracted parameters can be attributed to differences in the fabricated sample batches.

| Material | $\lambda_S$ (nm) | $\lambda_L$ (nm) | $\theta_{SH}$ | $\theta_{OH}$ |
|---|---|---|---|---|
| Pt | 1.0 to 1.6 | 1.2 to 1.6 | 0.0100 | $\ll 1$ |
| Ti | $\lambda_S \gg \lambda_L$ | 3.5 | $\ll 1$ | 0.1000 |
| Ge | $\lambda_S \gg \lambda_L$ | 3.8 | $\ll 1$ | -0.0290 |
| Au | - | - | 0.0004 | -0.0030 |